\documentclass[twocolumn,preprintnumbers,amsmath,amssymb]{revtex4}
\usepackage{graphics}
\usepackage{color}
\usepackage{epstopdf}
\usepackage{gensymb}
\begin{document}
\title{Ion microscopy based on laser-cooled cesium atoms}

\author{M.~Viteau$^1$, M.~Reveillard$^1$, L.~Kime$^1$, B.~Rasser$^1$ and P.~Sudraud$^1$, 
} 
\affiliation{$^1$ 
Orsay Physics, TESCAN Orsay, 95 Avenue des Monts Aur\'{e}liens - ZA Saint-Charles  - 13710 Fuveau, France}

%
%
\author{Y.~Bruneau$^2$, G. Khalili$^2$, P.~Pillet$^2$ and D.~Comparat$^2$} 

\affiliation{$^2$
Laboratoire Aim\'{e} Cotton, CNRS, Universit\'{e} Paris-Sud, ENS Cachan, B\^{a}t. 505, 91405 Orsay, France}

\author{I.~Guerri$^{3}$, A.~Fioretti$^{4,5,}$\footnote{corresponding author: andrea.fioretti@ino.it}, D.~Ciampini$^{3, 4, 5}$, M.~Allegrini$^{3, 4, 5}$ and F.~Fuso$^{3, 4, 5}$} 
\affiliation{$^{3}$
 Dipartimento di Fisica, Universit\`{a} di Pisa, Largo Pontecorvo 3, 56127 PISA, Italy}
\affiliation{$^{4}$
Istituto Nazionale di Ottica, INO-CNR, U.O.S.  "Adriano Gozzini", via Moruzzi 1, 56124 Pisa, Italy}
\affiliation{$^{5}$
Consorzio Nazionale Interuniversitario per le Scienze Fisiche della Materia, CNISM, Sezione di Pisa, 56127 Pisa, Italy}

\date{date: {\today}}
%

\begin{abstract}
{We demonstrate a  prototype of a Focused Ion Beam machine based on the ionization of a laser-cooled cesium beam adapted for imaging and modifying different surfaces in the few-tens nanometer range. Efficient atomic ionization is obtained by laser promoting ground-state atoms into a target excited Rydberg state, then field-ionizing them in an electric field gradient. The method allows obtaining ion currents up to 130~pA. Comparison with the standard direct photo-ionization of the atomic beam shows, in our conditions, a 40-times  larger ion yield. Preliminary imaging results at ion energies in the 1-5~keV range are obtained with a resolution around 40~nm, in the present version of the prototype. Our ion beam is expected to be extremely monochromatic, with an energy spread of the order of 1~eV,  offering great prospects for lithography, imaging and surface analysis.}
\end{abstract}
%
\maketitle

\section{Introduction}
\label{intro}

Charged particle beams of controlled energy and strong focusing are widely used tools in industry and science~\cite{mrs_2014}. State-of-the-art machines provide the possibility of modifying, analysing and imaging different objects and materials from the micro to the nano scale. As an example, a Focused Ion Beam (FIB) column can be combined with a Scanning Electron Microscope (SEM) to provide full control of nanofabrication or nanolithographic processes.  Ion energy can be varied typically in the 1-30 keV range, with an energy-dependent resolution attaining the nanometer range.
State-of-the-art FIBs are commercially available (see for example:~\cite{fei, tescan, op, zeiss}), based
mainly on plasma, liquid metal tip or helium ion sources for large, intermediate, and
low currents, respectively. Despite the very high technological level of the available machines, research of new ion sources allowing even higher resolution and a wider choice of atomic or molecular ions for new and demanding application is very active~\cite{Steele2014}. 

In the last decade, proposals~\cite{1028318, ref2005PhRvL..95p4801C, 2006JVSTB..24.2907H} and experimental realizations~\cite{Han08, ref2009PhRvL.102c4802R} of novel ion or electron sources based on the
ionization of laser-cooled atoms have been reported and have shown potential improvements with respect to standard sources, in particular a low energy-spread and a high resolution in the low-energy regime.
Due to the low temperatures associated with laser cooling, the ion
(or electron) beam originating from the cold sample has an extremely narrow angular
spread, i.e. a very low emittance. This means that ion or electron sources based on cold atom ionization
would be able to create at low acceleration energy very small focal spots with relatively strong currents.
Focused ion beams have been generated starting from both a magneto-optical trap (MOT)~\cite{Han08, ref2009PhRvL.102c4802R} and a slow and cold atomic beam~\cite{knuffman2013cold}. Their properties as nanoprobes have been demonstrated first with chromium~\cite{steele2010focused} and then with lithium~\cite{2011NJPh...13j3035K} ions, the latter reaching a 27~nm resolution at 2~keV energy and 1~pA current.
Very recently the lithium source has evolved into a complete system of scanning ion microscope able to work in the 500~eV - 5~keV low energy range~\cite{twedt2014}.
On the electron side, high-coherence electron bunches, obtained by ultrafast photo-ionization of cold atoms, have been demonstrated for single-shot electron diffraction studies~\cite{Engelen2013, McCulloch2013}.
It is worth mentioning that, in the 0.1 -5 fA current range, a laser-cooled atomic beam is being tested as a candidate for the controlled production of low-density ion beams~\cite{reza2015}. This could be of interest for ion implantation onto surfaces with nanometric precision aimed at engineering few atom devices~\cite{koenraad2011}.  

Here we present the prototypal realization of a complete low-energy Focused Ion Beam system based on
laser-cooled cesium atoms. The system represents the practical implementation of our original proposal~\cite{kime2013}. Atomic ionization is obtained 
either by laser promoting the neutral atoms into highly excited Rydberg states that are subsequently field-ionized, or by direct laser photo-ionization. 
The produced ions are coupled to a standard FIB column from Orsay Physics.
Once accelerated to energies in the range 1-5~keV, ions are used for imaging and surface modifications of different substrates. A current up to 130~pA is available, and a typical spatial resolution on the order of 40~nm at lower currents, largely independent of the ion acceleration in the explored range, is achieved in the produced images. 
Our results are a first step towards the realization of an industrial prototype of laser-cooled atom based FIB machines.

\section{Experimental setup}
\label{setup}

\begin{figure}[htbp]
\resizebox{0.5\textwidth}{!}{
   \begin{tabular}{c}
		 \includegraphics{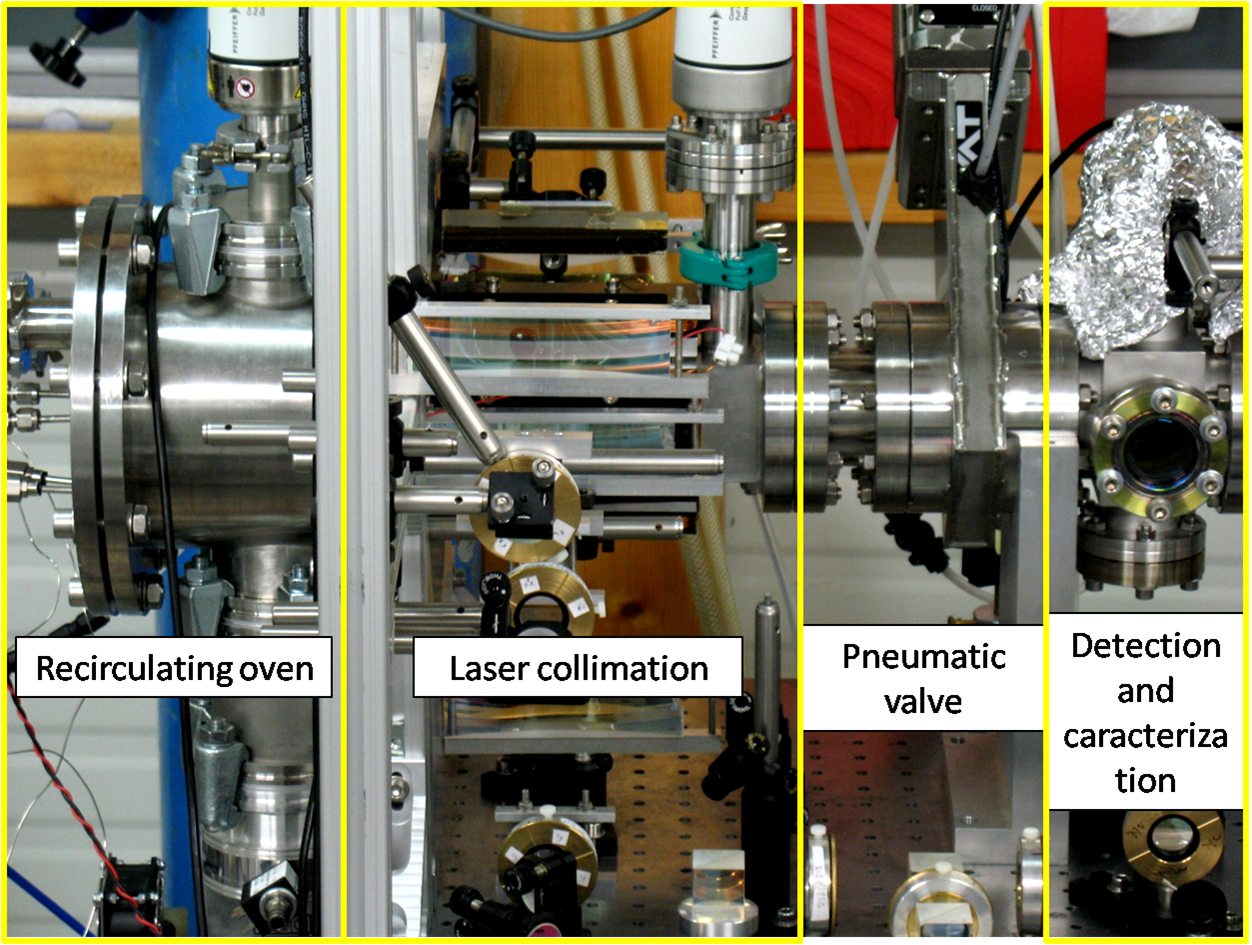} \\ 
		 \includegraphics{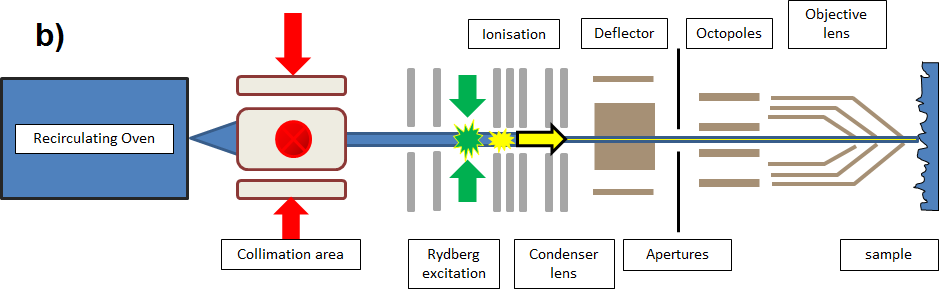} 
	 \end{tabular}
 	}
		\caption{(color online) a) Picture and b) sketch of the ultra-cold electron/ion source. An intense effusive atomic
beam is transversely cooled using laser cooling techniques. Electrons or ions (depending on electrode polarities) are
produced either by laser excitation to Rydberg states that are then field-ionized or by photo-ionization. The laser apparatus, the optical setup and the FIB column are not shown in the picture.
		}
\label{fig:setup}
\end{figure}

The experimental setup has been detailed in ref.~\cite{kime2013}. Briefly, it consists of a recirculating Cs oven outsourcing an effusive flux of cesium atoms, a laser-cooling region, an atomic excitation/ionization region, and finally a FIB column. A picture of part of the setup is shown in Fig.~\ref{fig:setup}. The oven delivers a typical flux of 1-2$\times 10^{13}$ atoms/s, with an average velocity of 200~m/s, for an oven temperature of 160~$^{\circ}$C. After the oven, these atoms are transversally laser-cooled by the interaction with a quasi-resonant laser light until their transverse temperature drops in the few hundreds microKelvin range~\cite{metcalf1994}. At this stage   the  atomic beam diameter is 4~mm and the estimated atom density is around $10^{10}$~cm$^{-3}$.  A few cm downstream atoms enter a  constant electric field region, where they are laser promoted to a highly-excited (Rydberg) state by interaction with two laser beams, and then a region of a few mm with of rapidly increasing electric field, where they are ionized at a specific field value. The ions are extracted and accelerated by the electric field towards the FIB column. At the end of the column they can be either focused on a sample mounted on a translation stage, or deflected onto a Faraday cup connected to a picoammeter to measure the current. 

Differently from most similar experiments~\cite{Han08, ref2009PhRvL.102c4802R, steele2010focused, knuffman2013cold} where photo-ionization is used, atomic ionization is obtained through Rydberg excitation. 
As detailed in ref.~\cite{kime2013}, the use of Rydberg atoms has two potential advantages compared to photo-ionization: a very large excitation efficiency and a low energy spread. The former quality is especially true for low principal quantum numbers $n$, and leads to a more efficient ionization of the atomic beam. The latter quality is largely independent of the excitation volume. Conversely, in the photo-ionization case, the energy spread is proportional to the product of the longitudinal length of the 
ionization region with the extraction electric field~\cite{ref2009PhRvL.102c4802R}. As a consequence, reducing the energy spread implies having limited ionization volumes, 
thus leading to small currents~\cite{2007JAP...102i4312V}. This is not the case for field ionization of Rydberg atoms, where the energy spread is essentially governed by the field gradient in the interaction region and by the characteristics of the chosen Rydberg state~\cite{kime2013}. The use of Rydberg atoms is therefore preferred because one can have in principle higher currents with the same energy spread and less laser power~\cite{kime2013}. 

The major technical drawback of the Rydberg method is the resonant character of the excitation and thus the necessity of a fine control of the laser frequency. In our case, Rydberg states with $n>$20 are produced by two-step excitation. A first diode laser (external cavity laser from Moglabs, with 60~mW power) is tuned and locked on the $6s_{1/2}{\rm (F=4)} \rightarrow 6p_{3/2}{\rm (F'=5)}$ transition at 852~nm, 
while a second frequency-doubled (SHG), external-cavity diode laser + MOPA amplifier (LEOS Solutions, maximum power 
300~mW at 510~nm), tunable in the 508-512~nm range (green), 
provides the Rydberg excitation on the $6p_{3/2}{\rm (F'=5)} \rightarrow ns_{1/2}/nd_{5/2}$ transitions.  This SHG laser can be tuned also above the ionization threshold, providing direct photo-ionization for comparison. The two excitation laser beams, both fiber-coupled, are sent orthogonally to one another and to the atomic beam. They are both focused to a $50 \pm 2$~$\mu$m waist and cross each other in their focal planes.  

\begin{figure}[htbp]
\centering
\resizebox{0.5\textwidth}{!}{
   \begin{tabular}{c}
		 \includegraphics{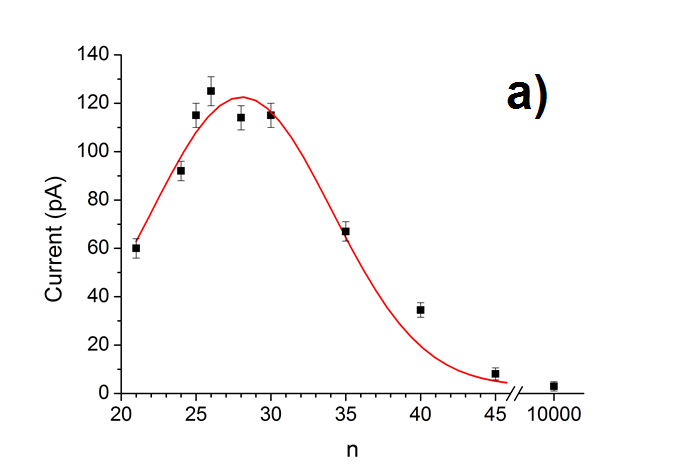}   \\
     \includegraphics{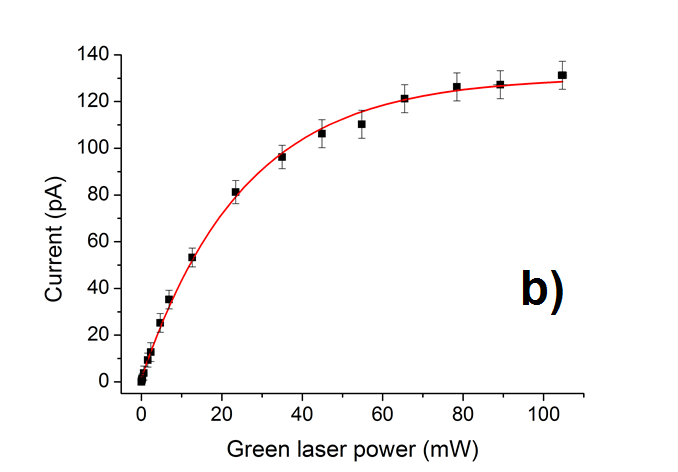}
	 \end{tabular}
}
\caption{a) Peak current in the FIB as a function of the principal quantum number $n$ of the excited Rydberg atom. Diaphragm diameter is 800~$\mu$m and green laser power 90~mW. The $n=10000$ point represents the photo-ionization above threshold. The solid line is a simple guide for the eyes. b) Peak current as a function of the green laser power for $n=26d_{5/2}$. The solid line is a simple guide for the eyes.}
\label{fig:exc_ryd}
\end{figure}

Once excited,  Rydberg atoms, which have the original longitudinal velocity of the effusive beam, enter the region of increasing electric field. Here, Rydberg atoms are field-ionized a few mm downstream, in a longitudinal position dependent on their $n$ level. Typical ionizing field values ($d$ states) are 3.5~kV/cm for $n=20$, 560~V/cm for $n=30$ and 160~V/cm for $n=40$~\cite{ioniz, gallagher1994}. The longitudinal profile of the field gradient is controlled by the potential of three closely-spaced electrodes, in a configuration similar to that described in ref.~\cite{kime2013}. The electric field also provides the necessary acceleration to extract the ions and send them into the FIB column.  By applying suitable voltages on the electrodes, different extraction energies are possible up to $5$~keV, currently limited by the electrical isolation of the feedthroughs.  With the present electrode design it is possible to obtain an ion signal for all levels $n>20$ up to the continuum of the direct photo-ionization.

The so-extracted ions are then directed into a standard FIB column (for a general description see Chap.~1 of ref.~\cite{lag04}), where the beams can be steered by two pairs of electrostatic deflectors. Subsequently, the ion beam can be apertured by remotely controlled diaphragms, from 10 to 800~$\mu$m diameter, which can improve the spatial resolution in the imaging owing to the decrease in the emittance. Such an improvement is inherently achieved at the expense of a decrease in the obtained current, meaning that longer integration times are necessary to obtain images with the same contrast.  At the end, an objective lens provides the final focusing into the sample while the scanning of the image is done by an octupole. In each scan, the integration time per pixel can be varied between 100~ns and 1~ms, thus defining the total dose of charge received by each pixel. 
Dedicated computer controls allow us to perform all the operations on the ion beam: adjusting the position and modifying the diameter of the diaphragm, varying the voltage of the many deflecting electrodes, defining the area of the sample to be scanned by the ion beam and  the integration time per pixel, thus determining the size and the duration of the scan. 
The output signal for the images is provided by secondary electrons emitted by the sample,  collected by a standard Everhart-Thornley detector mounted at an angle of $45\degree$  with respect to the sample surface.
The whole apparatus is kept under high vacuum conditions by 3 ionic and 1 turbo pumps. The vacuum is of the order of $8\times 10^{-7}$~mbar in the oven region, and 10$^{-7}$~mbar in the rest of the apparatus.

\section{Results and discussion}
\label{results}

We first investigated the maximum ion current produced by our system as a function of the level $n$ of the excited Rydberg atom. The results are shown in Fig.~\ref{fig:exc_ryd}a). As the excitation takes place in presence of an electric field, the atomic energy levels of the same $n$ and different angular momentum $l$ are mixed~\cite{gallagher1994} and many atomic lines are observed by varying the frequency of the green laser over a small interval. In absence of electric field one should expect only excitation of $s_{1/2}$  and $d_{3/2, 5/2}$ levels, because the starting level is the $6p_{3/2}$, but in fact many more lines are present 
in the non-zero electric field due to level mixing. The ion current in Fig.~\ref{fig:exc_ryd}a reports for each $n$ the value corresponding to the most intense line; the resulting graph has a bell-shaped trend. Towards lower $n$ numbers, where excitation 
is more favorable~\cite{excit, gallagher1994}, the current available with this method is limited by both the maximum electric field available for ionization and the lifetime of the Rydberg states, which becomes too short for excited atoms to reach the  field-ionization region before decaying. These are the reasons for the initial increase of the  current yield as a function of the level $n$ of the excited Rydberg atom. The maximum current is obtained for $n=26$, but the performances in terms of imaging capabilities are similar for $n$ values in the 24-30 range. For higher $n$ values, the current yield decreases again because of the decrease in the transition dipole moment of the $6p_{3/2} \rightarrow ns/nd$ transitions~\cite{excit, gallagher1994}. In our conditions, for a laser intensity around $I \sim 100$~W/cm$^2$, the current obtained  with the best Rydberg excitation is about a factor 40 larger than that obtained by above threshold photo-ionization. This represents a real advantage of our method.

Figure~\ref{fig:exc_ryd}b) shows the current as function of the green laser power for $n=26d_{5/2}$. For this value of $n$, the laser power is large enough to saturate the second step of the atomic excitation.
An ion current above 120~pA is typically produced for a green laser power larger than 60~mW, while currents up to 150~pA have been measured for higher oven temperatures and different laser configurations.

\begin{figure}[htbp]
\centering
\resizebox{0.4\textwidth}{!}{
     \includegraphics{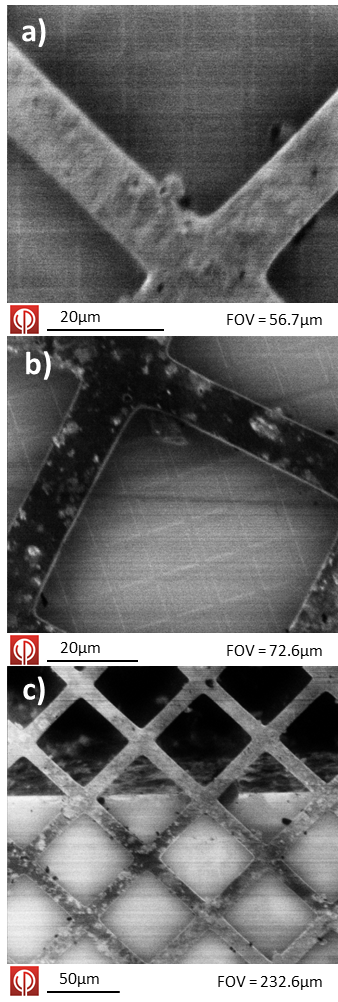}  
}
\caption{Images of a copper grid (50~$\mu$m pitch) on a silicon lithographed substrate (10~$\mu$m pitch) for different beam energies. a) 5~keV, 10~pA, 50~$\mu$s/point, 100~$\mu$m aperture, b) 3~keV, 10~pA, 50~$\mu$s/point, 100~$\mu$m aperture, c) 2~keV, 3~pA, 50~$\mu$s/point, 50~$\mu$m aperture. Calibrated field of view (FOV) are reported at the bottom of the images.}
\label{fig:diff_en}
\end{figure}

The performance of our complete FIB prototype based on the laser-cooled  cesium atomic source has been demonstrated through the analysis of different samples. 
An example of the imaging capabilities at low and intermediate beam energies (2-5~keV) is shown in Fig.~\ref{fig:diff_en}. Here, the specimen consists of a copper square grid placed over a lithographed silicon substrate. The copper grid is  a standard TEM mesh (nominal pitch 65~$\mu$m, bar width 10~$\mu$m), whereas the silicon substrate presents an array of square features (10~$\mu$m pitch, bar width 2~$\mu$m) produced by lithography. 
The shown images well elucidate the system performances at different ion beam energies.  The quality and resolution of the images are not affected by the acceleration voltage, at least down to 2~keV (Fig.~\ref{fig:diff_en}c). In conventional FIB systems, the possibility to operate at low ion beam energy is typically hampered by several limitations, including the chromatic aberration of the ion optics due to the non monochromatic energy distribution of the ion source~\cite{Jon08}. This is expected to be strongly reduced by ion sources based on laser-cooled atoms~\cite{ref2009PhRvL.102c4802R, kime2013}.

The shown images demonstrate also the capability of this prototype to detect and discriminate between different materials, owing to contrast mechanisms based on material properties and by possibly occurring charge accumulation at the surface. The latter phenomenon  is known to reduce secondary electron yield through the so-called passive voltage contrast~\cite{rosenkranz}, hence leading to a material dependent contrast mechanism.
The sensitivity to topography is also well demonstrated  by these images: defects on the copper bars are clearly visible, the surface features on the silicon substrate can be discerned even if they are due to very small height modulations.
Furthermore, a relatively large depth of field is achieved, which should be in the several hundreds $\mu$m range, given the estimated convergence angle of 1~mrad of the ions onto the sample. This can be seen, for instance, by looking at Fig.~\ref{fig:diff_en}c). Here, an irregular border of the substrate underlying the TEM mesh is imaged: small details can be discerned in both the mesh and the cleaved substrate.

\begin{figure}[htbp]
\resizebox{0.5\textwidth}{!}{
		 \includegraphics{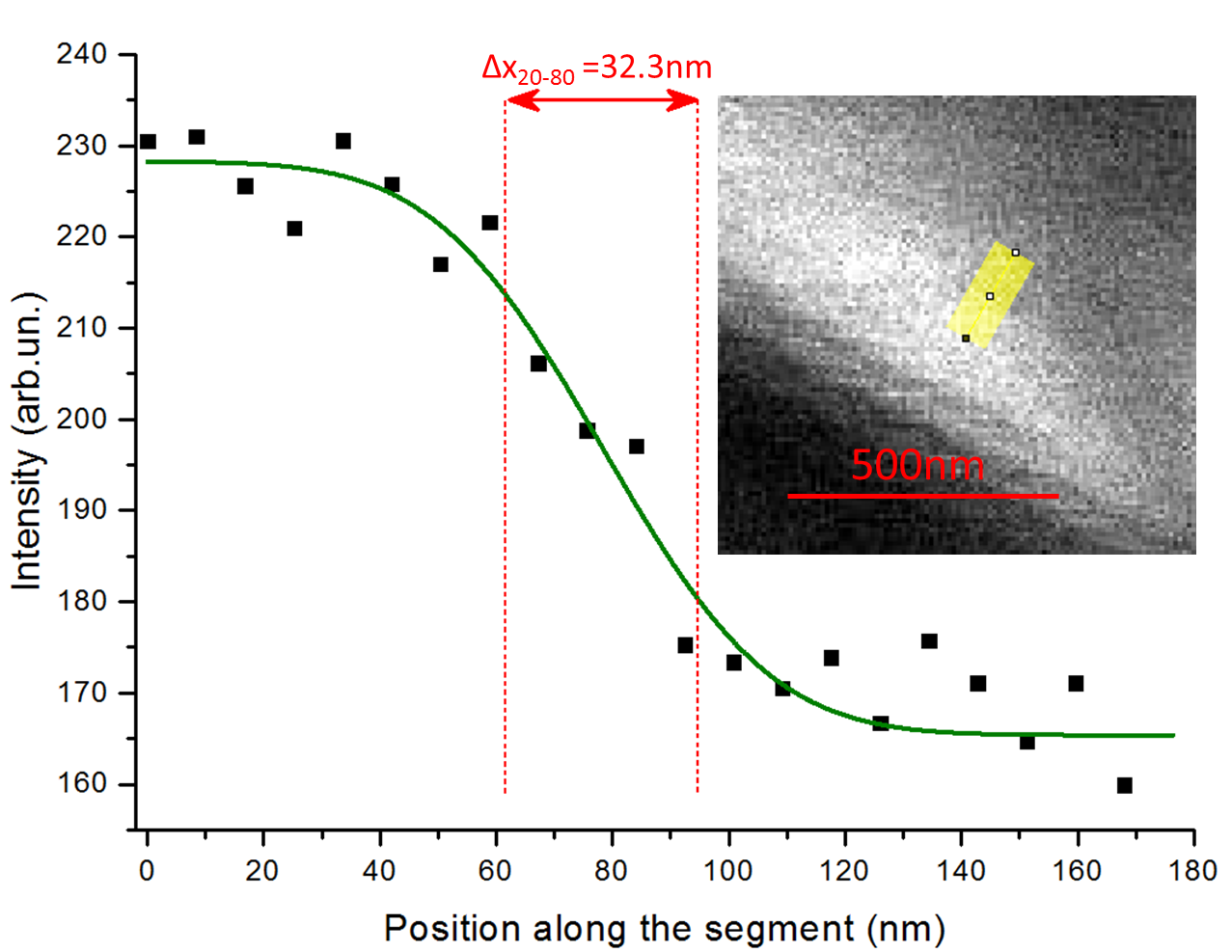} 
 	}
		\caption{(Color online) Cross section analysis of the sharp edge of the molybdenum surface of the Faraday cup shown in the inset (the thick yellow line represents the direction of the line profile). Experimental points are reported as dots. The green solid line is the result of a best-fit carried out with the ImageJ built-in function. The dashed vertical lines represent the 20-80\% variation interval of the best-fit function, leading to estimate the spatial resolution as $\Delta x_{20-80} = (32 \pm 10)$~nm. The image has been acquired at 2~keV acceleration energy, 20~$\mu$m aperture and 1~pA current.
}
\label{fig:reso}
\end{figure}

The spatial resolution of our FIB can be inferred by sectioning different images where the objects provide sharp edges. A typical example is shown in Fig.~\ref{fig:reso}, where the edge of a molybdenum feature of the Faraday cup is scanned. The analysis, carried out by using a conventional microscope software (ImageJ, NIH) and based on a best-fit to an error function of the line profile resulting from sectioning the image, suggests a spatial resolution, intended as the interval between the 20\% and the 80\% levels of the best-fit curve, below 40~nm. A better resolution ($\sim 20$~nm) has seldom been attained, but 40~nm is the reliable and reproducible value for day to day operation of the FIB prototype in the present stage of development. 


Such a spatial resolution turned out to be limited also by an irreversible misalignment between the laser-cooled atomic source, in particular the ionizing/condensing electrodes (see Fig.~\ref{fig:setup}b)), and the electrodes of the FIB column employed in the present experiment, leading to astigmatism of the beam. Moreover, the ion optics implemented in this initial stage of the column is conceived to operate with a point-like ion source such as the conventional Ga Liquid Metal Ion Source (LMIS). On the contrary, the transverse size of the  beam produced through ionization of the laser-cooled cesium beam is in the few hundred $\mu$m range, implying the occurrence of stray ion trajectories at the entrance of the column. Although this can be mitigated by aperturing the beam, 
a compromise must be found between the resulting aberration and the available current. We expect that a revised design of the ion optics, presently under development,  will suppress these aberration effects. 

\begin{figure}[htbp]
\centering
\resizebox{0.5\textwidth}{!}{
     \includegraphics{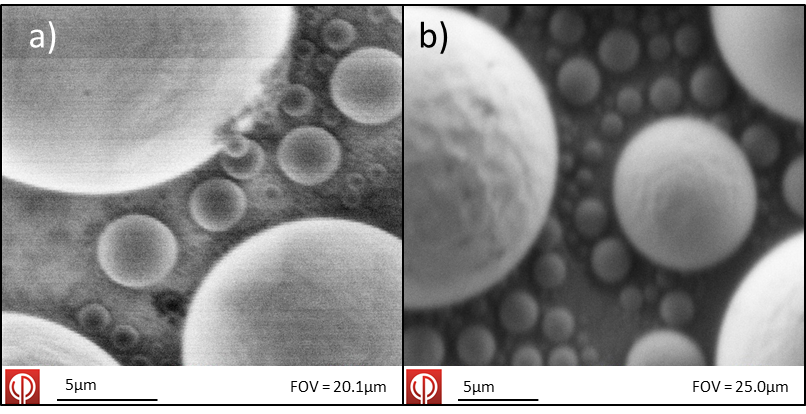} 
}
\caption{ Comparison between images acquired on a tin on carbon test sample with: a) our system at 5~keV ion beam energy and 7~pA current, and b) a commercial Ga LMIS FIB (Cobra-FIB by Orsay Physics, 5~keV ion beam energy, current 20~pA. Note the difference in contrast might be attributed to the change of the interactions between the substrate and the incident ions (Cs/Ga) or to the difference in mass. }
\label{fig:comparison}
\end{figure}

Despite of the mentioned limitations and as a further proof of the capabilities of our system to operate at intermediate voltages, we notice that the demonstrated spatial resolution around 40~nm is already comparable to what is  achieved with conventional, LMIS based, FIBs operated at intermediate ion beam energy. Figure~\ref{fig:comparison} shows images of test samples (tin droplets on carbon) acquired with our prototype and with a commercial FIB (Cobra-FIB by Orsay Physics), in panel a) and b), respectively. The overall quality of the images, depending on the spatial resolution and on the contrast, is comparable in both cases, or even better for our system. In fact, small tin droplets can be clearly discerned, as well as surface structures of the larger spheres. A few horizontal scars appear on the scan produced by our prototype (Fig.~\ref{fig:comparison}a), due to residual instabilities of the laser frequency. Such instabilities are expected to be removed in further implementations of the system, by making use of a more stable laser setup.

\begin{figure}[htbp]
\centering
\resizebox{0.5\textwidth}{!}{
     \includegraphics{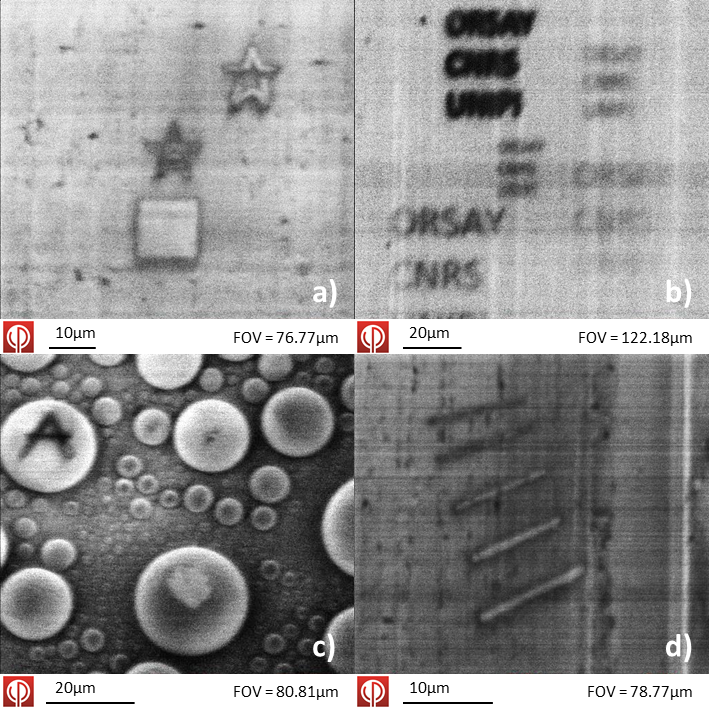} 
}
\caption{Examples of localized surface modifications produced by our prototype operated in the milling mode (ion beam energy 2~keV, current 2~pA). Stainless steel, a), b), and d), and tin on carbon, c), substrates are used.  The total exposure time to the cesium ion used for producing the straight line modifications shown in panel d) is (from top to bottom): 30~s, 60~s, 120~s, 240~s, 480~s. }
\label{fig:writing}
\end{figure}

The cesium FIB system has been tested also in terms of its capability of modifying a surface. To this aim, we have directed the focused ion beam onto selected surface regions of different samples, as in conventional FIB milling procedures. Subsequently, we have imaged the sample in order to analyze the effect of the ion/surface interaction. 
Some examples are reported in Fig.~\ref{fig:writing}, where stainless steel, a), b) d), and tin, c), surfaces are locally modified according to different motifs (geometrical shapes or characters), as enabled by the software controlling the FIB column. 

The low energy (2~keV) and most-of-all the very small current (2~pA) of the ion beam used in the tests are
expected not to produce remarkable removal of material. 
For instance, according to ref.~\cite{surfscirep}, removal rate of silicon substrates produced by Cs ions at 2 keV, normal incidence, is expected to be below one atom per incident ion.
On the other hand, at larger current yield, cesium ions sputtering properties are well known, making this element
widely used for negative ion sources based on surface bombardment~\cite{cookbook}. 
However, the shown examples clearly demonstrate the ability of our beam to induce surface modifications even in the low-current regime. In modified regions, we observe both dark and bright features, compared to the average level of the images, see, for instance, Fig.~\ref{fig:writing}a) and d). 
In the process, cesium ions are effectively milling the surface, e.g. sputtering material. This corresponds to the dark features, i.e. letters in Fig.~\ref{fig:writing}b) and c). At the same time, however, the cesium deposited onto the surface increases the production of the secondary electrons used for imaging.  In fact, cesium is known to modify the work function of the surfaces where it is deposited on. Therefore, after a certain exposure time, or dose of ions, brighter features can appear~\cite{lag04}.
This hypothesis is well confirmed by Fig.~\ref{fig:writing}d), where five straight segments are drawn on a stainless steel surface with different exposure times, as in the caption. The drawn features become brighter than the average level of the image for prolonged exposure, and, at the same time, their transverse width increases due to overexposure.
Evidence of the mechanism will be pursued in future experiments, where either tilting sample capabilities, or even a complete dual beam system, featuring an independent surface analysis by electron microscopy, will be implemented in the setup.

The quantitative analysis of the resolution performance of our system is beyond the scope of the present paper. 
Nonetheless, we can use the spatial resolution determined through the analysis presented in Fig.~\ref{fig:reso} to infer some relevant operating parameters of our ion source. The resolution of a microscope system is related to the focal spot size of the ion beam on the sample. Depending on the definition of resolution, the spot size can exceed the resolution by even an order of magnitude. Contributions to the spot size come from the demagnified source size in absence of aberrations, from the spherical and the chromatic aberrations and from the diffraction limit.
The spherical and chromatic aberration-limited spot sizes, according to ref.~\cite{Jon08}, are

\begin{eqnarray}
\label{aberr}
d_{\rm s}  & = & k_1 \frac{C_{\rm s}}{2} \alpha_{\rm i}^3  \\ 
d_{\rm c}  & = & k_2 C_{\rm c} \left(\frac{\Delta U_{\rm i}}{U_{\rm i}} \right) \alpha_{\rm i}  
\end{eqnarray}

where $k_1 = k_2 \simeq 0.5$ are numerical constants dependent on the fraction of the total current considered~\cite{barth1996}, $C_{\rm si}$ and $C_{\rm ci}$ are design parameters of the FIB column, $\alpha_{\rm i}$ is the angular beam divergence and $\Delta U_{\rm i}/U_{\rm i}$ the relative energy spread of the ions in the image plane. 

The aberration-free source size $d_{\rm ss}$ is given by (see~\cite{barth1996})
\begin{equation}
\label{dss}
d_{\rm ss} = \frac{2}{\pi} \frac{1}{\alpha_{\rm i}} \sqrt{\frac{I}{B_{\rm r}U_{\rm i}}} 
\end{equation}
where $I$ is the current and $B_{\rm r}$ the reduced brightness of the source.




The actual ion spot size can then be evaluated as (see~\cite{Jon08})

\begin{equation}
\label{d_total}
d \simeq \sqrt{d_{\rm ss}^2+d_{\rm s}^2+d_{\rm c}^2}, 
\end{equation}

where we have neglected the diffraction-limited spot size owing to the extremely small de Broglie wavelength of accelerated ions.

We estimated the brightness $B$ of our system  combining the experimental data with the results of a simulation carried out using the SIMION software. The simulation for the experimental parameters  (considering 50~\% of the current in the beam, a spot size around 330~nm, a current of 1~pA at 2~keV) gives: 
$\alpha_{\rm i} = 1$~mrad, $C_{\rm s}=1.5$~m and  $C_{\rm c}=0.5$~m. With  these values we obtain a reduced brightness of the beam  $B_{\rm r}=2.8 \times 10^5$~A~sr$^{-1}$~m$^{-2}$~eV$^{-1}$. 
Moreover, we can try to estimate the energy spread $\Delta U_{\rm i}$. By inserting these values into Eqs~(1-3) we obtain that the estimated ion spot size $d$ is largely dominated by the chromatic aberration, being  the aberration-free term $d_{\rm ss} \sim 27$~nm and the spherical aberration-limited one $d_{\rm s} \sim 0.4$~nm. Thus, we would obtain an energy spread  of the order of $\Delta U_{\rm i} \sim 2$~eV for 2~keV ions. This is roughly compatible with the estimated energy spread for the field ionization of the selected ($n \sim 30$) Rydberg level, see Table III in ref.~\cite{kime2013}.


Even if dedicated experimental investigations are necessary to clarify the role of the different parameters that actually determine the performances of our FIB, we can estimate that using other ``exceptional'' Rydberg states (see ref.~\cite{kime2013}) may lead to a smaller energy dispersion and thus to spot sizes no longer limited by chromatic aberrations.




\section{Conclusions and outlook}
\label{conclusions}

In conclusion, we have demonstrated the potential of a new source for FIB columns using a laser-collimated cesium atomic beam, ionized to produce a low energy dispersion ionic beam. The ionization is done by a two-step process, with first an excitation to a Rydberg state and a further ionization by an electric field. By coupling this ion beam to a standard FIB column we show its ability to produce images with a spatial resolution below 40~nm at low energies (2-5~keV). We also demonstrate modifications of the surface interpreted as due to a combination of surface milling and cesium deposition.

A new generation of electrostatic optics is under realisation to overcome the present intrumental limits. This will demonstrate the ultimate possibilities of this FIB prototype, based on the new approach of field ionization of Rydberg atoms, excited from a laser-cooled atomic beam. Moreover, 
a laser-cooling compression phase, located between the collimation and the excitation/ionization regions, would be important to increase the neutral atomic density prior to ionization~\cite{metcalf1994}. This eventual implementation is expected to increase the current and possibly the brightness of the source. The consequent reduction of the atomic beam radius would also reduce laser power requirements, as lasers could be focused more tightly.   At this point, due to the expected increase in the current density, problems arising from Coulomb forces among particles will come into play~\cite{kime2013}, degrading the emittance and thus the brightness. Nevertheless, extended simulations of the expected behavior of a laser intensified thermal beam used as an ion source for a FIB have been recently performed in Eindhoven, showing that nanometer size spots can be attained with a 30~keV beam of Rubidium ions up to currents of a few pA, outperforming a LMIS based FIB in terms of minimum  attainable spot size~\cite{tenHaaf2014, Wouters2014}.

{\bf Acknowledgments}: 
We thank C.~Colliex, A.~Gloter and F.~Robicheaux for helpful discussions. We thank N.~Porfido and J.~Gurian for contributions in the early stage of the experiment, E.~Andreoni and N.~Puccini for technical support.
The research leading to these results has received funding from Institut Francilien de Recherche sur les Atomes Froids
(IFRAF), f\'ed\'eration de recherche Lumi\`ere Mati\`ere (LUMAT), and the European Union.  
D.C., F.F. and P.S. acknowledge gratefully the support of the European Union
Seventh Framework Program FP7/2007-2013 under Grant Agreement No. 251391
MC-IAPP COLDBEAMS. I.G. and M.V. have been Research Fellows hired under this
program. D.C acknowledge also funding from the European
Research Council under the ERC grant agreement n.~277762 COLDNANO.
A.F. has been supported by the {\it Triangle de la Physique}
under Contracts n.~2007-n.74T and n.~2009-035T GULFSTREAM.

This article is in memory of Pierre Sudraud, outstanding inventor and entrepreneur.

\bibliographystyle{unsrt}
\bibliography{2014_bibli_global_ions}

\begin{thebibliography}{10}

\bibitem{mrs_2014}
Nabil Bassim, Keana Scott, and Lucille~A. Giannuzzi, editors.
\newblock {\em Focused Ion Beam Technology and Applications}.
\newblock MRS Bullettin, Vol. 39, 2014.

\bibitem{fei}
http://www.fei.com/products/.

\bibitem{tescan}
http://www.tescan.com/en.

\bibitem{op}
http://www.orsayphysics.com/.

\bibitem{zeiss}
http://www.zeiss.com/microscopy/en\_de/products.html.

\bibitem{Steele2014}
Noel~S. Smith, John~A. Notte, and Adam~V. Steele.
\newblock Advances in source technology for focused ion beam instruments.
\newblock {\em MRS Bulletin}, 39:329--335, 2014.

\bibitem{1028318}
B.~G. {Freinkman}, A.~V. {Eletskii}, and S.~I. {Zaitsev}.
\newblock A proposed laser source of ions for nanotechnology.
\newblock {\em Microelectron. Eng.}, 73-74(1):139--143, 2004.

\bibitem{ref2005PhRvL..95p4801C}
B~J Claessens, S~B van~der Geer, G~Taban, E~J Vredenbregt, and O~J Luiten.
\newblock Ultracold electron source.
\newblock {\em Physical Review Letters}, 95(16):164801, 2005.

\bibitem{2006JVSTB..24.2907H}
J.~L. {Hanssen}, E.~A. {Dakin}, J.~J. {McClelland}, and M.~{Jacka}.
\newblock Using laser-cooled atoms as a focused ion beam source.
\newblock {\em Journal of Vacuum Science and Technology B: Microelectronics and
  Nanometer Structures}, 24:2907, 2006.

\bibitem{Han08}
J.~L. {Hanssen}, S.~B. {Hill}, J.~{Orloff}, and J.~J. {McClelland}.
\newblock Magneto-optical-trap-based, high brightness ion source for use as a
  nanoscale probe.
\newblock {\em Nano Letters}, 8:2844, 2008.

\bibitem{ref2009PhRvL.102c4802R}
M.P. {Reijnders}, P.A. {van Kruisbergen}, G.~{Taban}, S.B. {van der Geer}, P.H.
  {Mutsaers}, E.J. {Vredenbregt}, and O.J. {Luiten}.
\newblock Low-energy-spread ion bunches from a trapped atomic gas.
\newblock {\em Physical Review Letters}, 102(3):034802, 2009.

\bibitem{knuffman2013cold}
B~Knuffman, AV~Steele, and JJ~McClelland.
\newblock Cold atomic beam ion source for focused ion beam applications.
\newblock {\em Journal of Applied Physics}, 114(4):044303--044303, 2013.

\bibitem{steele2010focused}
AV~Steele, B.~Knuffman, JJ~McClelland, and J.~Orloff.
\newblock Focused chromium ion beam.
\newblock {\em Journal of Vacuum Science \& Technology B: Microelectronics and
  Nanometer Structures}, 28:C6F1, 2010.

\bibitem{2011NJPh...13j3035K}
B.~{Knuffman}, A.~V. {Steele}, J.~{Orloff}, and J.~J. {McClelland}.
\newblock {Nanoscale focused ion beam from laser-cooled lithium atoms}.
\newblock {\em New Journal of Physics}, 13(10):103035, October 2011.

\bibitem{twedt2014}
K.A. {Twedt}, L.~{Chen}, and J.J. {McClelland}.
\newblock Scanning ion microscopy with low energy lithium ions.
\newblock {\em Ultramicroscopy}, 142:24--31, 2014.

\bibitem{Engelen2013}
W.J. {Engelen}, M.A. {van der Heijden}, D.J. {Bakker}, E.J.D. {Vredenbregt},
  and O.J. {Luiten}.
\newblock High-coherence electron bunches produced by femtosecond
  photoionization.
\newblock {\em NATURE COMMUNICATIONS}, 4:1693, 2013.

\bibitem{McCulloch2013}
A.J. {McCulloch}, D.V. {Sheludko}, M.~{Junker}, and R.E. {Scholten}.
\newblock High-coherence picosecond electron bunches from cold atoms.
\newblock {\em NATURE COMMUNICATIONS}, 4:1962, 2013.

\bibitem{reza2015}
G.~Shayeganrad, A.~Fioretti, I.~Guerri, F.~Tantussi, D.~Ciampini, M.~Allegrini,
  M.~Viteau, and F.~Fuso.
\newblock Low-energy ions from laser-cooled atoms.
\newblock {\em submitted to Phys Rev. Appl.}, 2015.

\bibitem{koenraad2011}
P.M. Koenraad and M.E. Flatte.
\newblock Single dopants in semiconductors.
\newblock {\em Nature Materials}, 10:91, 2011.

\bibitem{kime2013}
L.~{Kime}, A.~{Fioretti}, Y.~{Bruneau}, N.~{Porfido}, F.~{Fuso}, M.~{Viteau},
  G.~{Khalili}, N.~{Santi\'c}, A.~{Gloter}, B.~{Rasser}, P.~{Sudraud},
  P.~{Pillet}, and D.~{Comparat}.
\newblock High-flux monochromatic ion and electron beams based on laser-cooled
  atoms.
\newblock {\em PHYSICAL REVIEW A}, 88:033424, 2013.

\bibitem{metcalf1994}
H.~Metcalf and P.~Van der Straten.
\newblock Cooling and trapping of neutral atoms.
\newblock {\em Phys. Rep.}, 244:203, 1994.

\bibitem{2007JAP...102i4312V}
S.~B. {van der Geer}, M.~P. {Reijnders}, M.~J. {de Loos}, E.~J.~D.
  {Vredenbregt}, P.~H.~A. {Mutsaers}, and O.~J. {Luiten}.
\newblock Simulated performance of an ultracold ion source.
\newblock {\em Journal of Applied Physics}, 102:094312, 2007.

\bibitem{ioniz}
Rydberg states ionize at the electric field of $\approx 400 \times (
  30/n^{*})^4$ v/cm for alkali-metal atoms, where $n^{*}$ is the effective
  principle quantum number.

\bibitem{gallagher1994}
Thomas~F. Gallagher.
\newblock {\em Rydberg Atoms}.
\newblock Cambridge University Press, Cambridge, 1994.

\bibitem{lag04}
L.A. Giannuzzi and F.A. Stevie, editors.
\newblock {\em Introduction to Focused Ion Beams: Instrumentation, Theory,
  Techniques and Practice}.
\newblock Springer, 2005.

\bibitem{excit}
The optical excitation cross-section of a resolved rydberg state has a $n^{-3}$
  dependence.

\bibitem{Jon08}
Jon Orloff.
\newblock {\em Handbook of Charged Particle Optics}.
\newblock CRC Press, 2008.

\bibitem{rosenkranz}
R.~Rosenkranz.
\newblock Failure localization with active and passive voltage contrast in fib
  and sem.
\newblock {\em J. Mater. Sci.: Mater. Electron.}, 32:1523--1535, 2011.

\bibitem{surfscirep}
K.~Wittmaack.
\newblock Unravelling the secrets of cs controlled secondary ion formation:
  Evidence of the dominance of site specific surface chemistry, alloying and
  ionic bonding.
\newblock {\em Surface Science Reports}, 68:108 -- 230, 2013.

\bibitem{cookbook}
Roy Middleton.
\newblock {\em A negative-Ion cookbook}.
\newblock 1990.

\bibitem{barth1996}
J.~E. Barth and P.~Kruit.
\newblock Addition of different contributions to the charged particule probe
  size.
\newblock {\em Optik}, 101:101, 1996.

\bibitem{tenHaaf2014}
G.~ten Haaf, S.H.W. Wouters, S~B van~der Geer, E.J.D. Vredenbregt, and P.H.A.
  Mutsaers.
\newblock Performance predictions of a focused ion beam from a laser cooled and
  compressed atomic beam.
\newblock {\em Journal of Applied Physics}, 116:244301, 2014.

\bibitem{Wouters2014}
S.H.W. Wouters, G.~ten Haaf, R.P.M.J.W. Notermans, N.~Debernardi, P.H.A.
  Mutsaers, O.J. Luiten, and E.J.D. Vredenbregt.
\newblock Performance predictions for a laser-intensified thermal beam for use
  in high-resolution focused-ion-beam instruments.
\newblock {\em PHYSICAL REVIEW A}, 90:063817, 2014.

\end{thebibliography}

\end{document}